\documentstyle[preprint,aps]{revtex}

\tightenlines

\begin{document}


\draft

\title{\bf 
Field Dependence of the Magnetization for the spin-ladder
material Cu$_2$(C$_5$H$_{12}$N$_2$)$_2$Cl$_4$}

\author{Zheng Weihong$^1$\cite{byline1}, Rajiv R.P. Singh$^2$\cite{byline2},
and J. Oitmaa$^1$\cite{byline3}} 
\address{${}^1$School of Physics,                                              
The University of New South Wales,                                   
Sydney, NSW 2052, Australia.\\                      
${}^2$Department of Physics,                                              
University of California,                                   
Davis, CA 95616, USA.}                      

\date{\today}

\maketitle 

\begin{abstract}
We have developed a series expansion method for calculating the
uniform magnetization, $M$, as a function of the applied field $h$
for Heisenberg systems at $T=0$. The method involves introducing
Ising anisotropy along the z-axis together with an applied uniform
field along the $x$-axis. On extrapolation to the isotropic limit,
one recovers the magnetization for the Heisenberg system with an applied
field along the x-axis. This method circumvents the difficulties
in developing perturbation theory associated with the commuting
nature of the uniform field. The results developed for two-chain
ladders appropriate for the
material Cu$_2$(C$_5$H$_{12}$N$_2$)$_2$Cl$_4$ are in good agreement
with the experimental data. In addition, uniform susceptibility
is calculated by high temperature expansions and also compared
with the experimental data.
\end{abstract}                                                              


\narrowtext
Recently several spin-ladder materials have been discovered,
which exhibit properties characteristic of one-dimensional systems
\cite{tak92,dag95}.
The material Cu$_2$(C$_5$H$_{12}$N$_2$)$_2$Cl$_4$ represents a
particularly simple Heisenberg spin-system from a theoretical
point of view as it forms \cite{Levy} a spin-ladder with
the inter-chain coupling, $J_\perp$ much larger than the
intra-chain coupling, $J_{{}_{/\!/}}$, making it ideally suited for
perturbation theory and for testing out quantitative theoretical
calculations. For this system, the nuclear relaxation rate
$1/T_1$ along with the Knight shift and field dependent
magnetization have been measured by Levy et al. Here, we
would like to study this system by series expansion method
and in particular try out a new method for calculating
field dependent properties of the Heisenberg spin systems.

In recent years series expansion methods have played a significant
role in studying magnetic properties of quantum spin models and
real materials \cite{he90,gel90}. In addition to high temperature expansions,
a variety of $T=0$ perturbation expansions have been developed
to study quantum spin Hamiltonians. Important technical
advances have been made in calculating the excitation
spectra by these methods \cite{gelfand}.

One general problem that, to our knowledge, has not been considered
before is related to dependence on properties associated
with a commuting variable. Example of such variables are
chemical potential and uniform magnetic field. The difficulty
posed by such variables is that at $T=0$ different values
of magnetization or density lead to different sectors of
the Hilbert space, which are not connected to each other.
Thus, naively, one would need to consider all different
sectors of the Hilbert space to find the global minimum
and hence determine the dependence on these fields.
This is clearly a formidable task.

Here we circumvent this difficulty by explicitly
breaking the symmetry at intermediate stages of the
calculation. For the Heisenberg-Ising
model in a transverse field, the field term no longer
commutes with the Hamiltonian. By series extrapolation,
we recover the properties of the Heisenberg limit, where
the field term does commute with the Hamiltonian. It is assumed that
by the limiting procedure, implicit in the series analysis,
the global ground-state with
respect to all possible magnetizations is obtained in
the Heisenberg limit. Comparison with experimental data
confirms that this is a valid procedure for calculating 
the field dependent properties.

We begin with the spin-ladder Hamiltonian in an external field
\begin{equation}
H=H_l+{\bf h}\cdot\sum_i {\bf S_i}
\end{equation}
where $H_l$ represents the spin-ladder Hamiltonian in zero field.
High temperature expansions for the system with
$J_\perp=J_{{}_{/\!/}}$, as well as $T=0$ Ising and Dimer
expansions for this model for $h=0$ have been discussed earlier \cite{osz}.

Here, in order to compare with the  experimental data for
Cu$_2$(C$_5$H$_{12}$N$_2$)$_2$Cl$_4$,
and determine the exchange constants for
the system, we calculate the high temperature series
for uniform susceptibility $\chi(T)$ with arbitrary
$y_2 \equiv J_{{}_{/\!/}}/J_\perp$. The series coefficients 
are given in table 1.
To make the comparison easier,
we take $J_\perp/J_{{}_{/\!/}}=5$, a value inferred 
earlier by Levy et al. \cite{Levy}.
Since the susceptibility is presented in arbitrary units, we are left
with two fitting parameters: an overall magnitude of $\chi$ is
set by a free parameter, whereas the fit of the temperature scale is
controlled by the exchange constant $J_\perp$.
We find that the best fit is obtained with $J_\perp=13.96$K.
Comparison with experimental data is shown in Fig. 1
and verifies previous assertions that the above model 
with $J_\perp/J_{{}_{/\!/}}=5$ is 
appropriate for the material under consideration.

In this paper we are primarily interested in $T=0$ field
dependent magnetization. The properties in zero field
including the spin-gap is most accurately calculated by
the dimer expansions. The gap is estimated to be 
$\Delta=4.10855 J_{{}_{/\!/}}$.
However, as discussed earlier,
this expansion is not suitable for calculating
the field dependent properties. Instead, we consider
the following model which depends on the parameter
$\lambda$,
\begin{eqnarray}
H &=& H_0 + \lambda V  \\
H_0 &= & J_{{}_{/\!/}} \sum_{i,l=1}^{l=2} S_{l,i}^z S_{l,i+1}^z +
  J_\perp \sum_{i} S_{1,i}^z S_{2,i}^z   + t \sum_{l,i} (-1)^{l+i} S_{l,i}^z \nonumber \\
\nonumber \\
V &= & J_{{}_{/\!/}} \sum_{i,l=1}^{l=2} 
( S_{l,i}^x S_{l,i+1}^x + S_{l,i}^y S_{l,i+1}^y ) +
J_\perp \sum_{i} ( S_{1,i}^x S_{2,i}^x + S_{1,i}^y S_{2,i}^y ) 
- t \sum_{l,i} (-1)^{l+i} S_{l,i}^z + h \sum_{l,i} S_{l,i}^x
\nonumber 
\end{eqnarray}   
where ${\bf S}_{l,i}$ denotes the S=1/2 spin at the $i$th site of the $l$th
chain. 
Notice that the staggered field terms (the terms with coefficient $t$)
are simply added to improve convergence, and we adjust the
coefficient $t$ to get the smoothest terms in series with
a typical value being $t =2 J_\perp$ . For $\lambda=1$
the model reduces to the spin-ladder Heisenberg
model in a uniform field. However, for $\lambda<1$,
the field term does not commute with the Hamiltonian
and thus mixes states with different magnetization.
In practice, the only complication this brings is that the
dimensionality of Hilbert space for a given graph that
needs to be considered is increased.

We have obtained the expansions for uniform magnetization $M$,
the expectation value of $1/N \sum_{l,i} S_{l,i}^x$ in the ground state,
 up to order $\lambda^{14}$ for
a range of field values (the series  coefficients
for several $h/J_{{}_{/\!/}}$ values
are presented in Table 2), the calculations involved
a list of 2767 linked cluster of up to 14 sites. 
The resulting series are
extrapolated to the Heisenberg point by Pad\'{e} and
differential approximants \cite{gut}. The resulting plots
for the magnetization are shown in Fig. 2. It
is encouraging that the magnetization appears to
go to zero at the onset field, $h_{c1}=4.10855 J_{{}_{/\!/}}$
known accurately from the gap calculated
from the dimer expansions. Although, the dependence 
of the magnetization on the field
is linear for most of the field values, near the onset field $h_{c1}$, 
it can be fit to a power law behavior
\begin{eqnarray}
M =c [(h-h_{c1})/J_{{}_{/\!/}}]^{\beta}.
\end{eqnarray}
Using the accurately known $h_{c1}$ and performing two-point fits leads
to the estimates $\beta=0.55(10)$ and $c=0.18(3)$, 
consistent with theoretical expectations\cite{giam}. 
Near the upper critical field $h_{c2}$, where the magnetization saturates
to the maximum value of $1/2$,
our results become less accurate, making it difficult to obtain
any critical exponents. The ratio $h_{c2}/h_{c1}$
is $\approx 1.8$ in our calculations, which is close to the
experimental value of $1.7$. The full plots are in
reasonable agreement with the experimental data.

Conclusions: In this paper, we have developed a series expansion
method for calculating the field dependent properties of the 
Heisenberg model. The method is shown to work well for the
spin-ladder system and provides a good description for the
material Cu$_2$(C$_5$H$_{12}$N$_2$)$_2$Cl$_4$. In future, 
we hope to use this method to study the field dependent
properties of the Heisenberg models in other geometries.

\acknowledgments
This work in UNSW forms part of a research project supported by a grant 
from the Australian Research Council. R. R. P. S. is supported
in part by the National Science Foundation through grant number
DMR-9318537. We would like to thank L.P. Levy for sharing with us the 
experimental data.


\begin{figure}
\caption{Comparison of the calculated susceptibility $\chi$ for the system with
$J_{{}_{/\!/}} /J_\perp=0.2$ (shown as the solid lines) with the experimental data 
(shown as open symbols) of Levy for the
material Cu$_2$(C$_5$H$_{12}$N$_2$)$_2$Cl$_4$.}
\label{fig:fig1}
\end{figure}

\begin{figure}
\caption{Uniform magnetization $M$ for the system with
$J_{{}_{/\!/}} /J_\perp=0.2$ as a function of applied field $h/J_{{}_{/\!/}}$
obtained by Ising expansion, the solid line is the power law behavior
$M= 0.18[(h-h_{c1})/J_{{}_{/\!/}}]^{0.55}$ estimated by the two-point fits
to the region near $h_{c1}$.}
\label{fig:fig2}
\end{figure}

\setdec 0.000000000000
\begin{table}
\squeezetable
\caption{Series coefficients for high-temperature expansion of the
uniform susceptibility $ T \chi (T) = \sum_{n,m} a_{n,m} (y_2)^{n} \beta^m$
of the system with  $y_2=J_{{}_{/\!/}} /J_\perp$, and $\beta = J_\perp/T$.
Nonzero coefficients $a_{n,m}$
up to order $m=9$ are listed.}\label{tabchi}
\begin{tabular}{rr|rr|rr|rr}
\multicolumn{1}{c}{(n,m)} &\multicolumn{1}{c|}{$a_{n,m}$}
&\multicolumn{1}{c}{(n,m)} &\multicolumn{1}{c|}{$a_{n,m}$}
&\multicolumn{1}{c}{(n,m)} &\multicolumn{1}{c|}{$a_{n,m}$}
&\multicolumn{1}{c}{(n,m)} &\multicolumn{1}{c}{$a_{n,m}$} \\
\hline
( 0, 0) &\dec  2.500000000$\times 10^{-1}$ &( 0, 5) &\dec  2.115885417$\times 10^{-4}$
 &( 1, 7) &\dec  2.556694878$\times 10^{-4}$ &( 7, 8) &\dec $-$3.694080171$\times 10^{-4}$ \\
( 0, 1) &\dec $-$6.250000000$\times 10^{-2}$ &( 1, 5) &\dec $-$1.790364583$\times 10^{-3}$
 &( 2, 7) &\dec $-$5.249023438$\times 10^{-4}$ &( 8, 8) &\dec  2.766200474$\times 10^{-4}$ \\
( 1, 1) &\dec $-$1.250000000$\times 10^{-1}$ &( 2, 5) &\dec  3.255208333$\times 10^{-3}$
 &( 3, 7) &\dec  6.212022569$\times 10^{-4}$ &( 0, 9) &\dec  3.899804709$\times 10^{-6}$ \\
( 0, 2) &\dec $-$1.562500000$\times 10^{-2}$ &( 3, 5) &\dec $-$2.604166667$\times 10^{-3}$
 &( 4, 7) &\dec $-$4.855685764$\times 10^{-4}$ &( 1, 9) &\dec $-$2.440800742$\times 10^{-5}$ \\
( 1, 2) &\dec  6.250000000$\times 10^{-2}$ &( 4, 5) &\dec  9.765625000$\times 10^{-4}$
 &( 5, 7) &\dec  5.045572917$\times 10^{-4}$ &( 2, 9) &\dec  4.863436260$\times 10^{-5}$ \\
( 0, 3) &\dec  1.302083333$\times 10^{-3}$ &( 5, 5) &\dec $-$1.367187500$\times 10^{-3}$
 &( 6, 7) &\dec  9.087456597$\times 10^{-5}$ &( 3, 9) &\dec $-$6.667041274$\times 10^{-5}$ \\
( 1, 3) &\dec  7.812500000$\times 10^{-3}$ &( 0, 6) &\dec $-$1.044379340$\times 10^{-4}$
 &( 7, 7) &\dec  6.200396825$\times 10^{-5}$ &( 4, 9) &\dec  1.052977547$\times 10^{-4}$ \\
( 2, 3) &\dec $-$1.562500000$\times 10^{-2}$ &( 1, 6) &\dec  2.766927083$\times 10^{-4}$
 &( 0, 8) &\dec  1.513768756$\times 10^{-6}$ &( 5, 9) &\dec $-$1.893664163$\times 10^{-4}$ \\
( 3, 3) &\dec  1.041666667$\times 10^{-2}$ &( 2, 6) &\dec $-$3.255208333$\times 10^{-5}$
 &( 1, 8) &\dec  1.850430928$\times 10^{-5}$ &( 6, 9) &\dec  1.067227157$\times 10^{-4}$ \\
( 0, 4) &\dec  1.627604167$\times 10^{-3}$ &( 3, 6) &\dec  1.736111111$\times 10^{-4}$
 &( 2, 8) &\dec $-$1.121399895$\times 10^{-4}$ &( 7, 9) &\dec $-$6.055075025$\times 10^{-5}$ \\
( 1, 4) &\dec $-$5.208333333$\times 10^{-3}$ &( 4, 6) &\dec $-$1.155598958$\times 10^{-3}$
 &( 3, 8) &\dec  1.617916047$\times 10^{-4}$ &( 8, 9) &\dec $-$1.049223400$\times 10^{-4}$ \\
( 2, 4) &\dec  3.906250000$\times 10^{-3}$ &( 5, 6) &\dec  1.497395833$\times 10^{-3}$
 &( 4, 8) &\dec  5.907331194$\times 10^{-5}$ &( 9, 9) &\dec  5.028403415$\times 10^{-5}$ \\
( 3, 4) &\dec $-$5.208333333$\times 10^{-3}$ &( 6, 6) &\dec $-$1.082356771$\times 10^{-3}$
 &( 5, 8) &\dec $-$2.031598772$\times 10^{-4}$  \\
( 4, 4) &\dec  3.255208333$\times 10^{-3}$ &( 0, 7) &\dec $-$3.986661396$\times 10^{-5}$
 &( 6, 8) &\dec  4.343184214$\times 10^{-4}$  \\
\end{tabular}
\end{table}

\setdec 0.00000000000000
\begin{table}
\squeezetable
\caption{Series coefficients of the
uniform magnetization $M = \sum_{i} a_{i} \lambda^i $
of the system with  $J_{{}_{/\!/}}/J_\perp=0.2$ and staggered field $t=2 J_\perp$.
Nonzero coefficients $a_{i}$
up to order $i=14$ are listed.}\label{tabM}
\begin{tabular}{rcccc}
\multicolumn{1}{c}{$i$} &\multicolumn{1}{c}{$a_i$ for $h/J_{{}_{/\!/}}=4$}
&\multicolumn{1}{c}{$a_i$ for $h/J_{{}_{/\!/}}=4.5$} 
&\multicolumn{1}{c}{$a_i$ for $h/J_{{}_{/\!/}}=5$}
&\multicolumn{1}{c}{$a_i$ for $h/J_{{}_{/\!/}}=5.5$} \\
\hline
  1 &\dec $-$2.35294117647$\times 10^{-1}$ &\dec $-$2.64705882353$\times 10^{-1}$
 &\dec $-$2.94117647059$\times 10^{-1}$ &\dec $-$3.23529411765$\times 10^{-1}$ \\
  2 &\dec $-$1.09573241061$\times 10^{-2}$ &\dec $-$1.23269896194$\times 10^{-2}$
 &\dec $-$1.36966551326$\times 10^{-2}$ &\dec $-$1.50663206459$\times 10^{-2}$ \\
  3 &\dec   9.14025678774$\times 10^{-2}$ &\dec   1.06396227962$\times 10^{-1}$
 &\dec   1.22649301848$\times 10^{-1}$ &\dec   1.40301724400$\times 10^{-1}$ \\
  4 &\dec   6.20129611723$\times 10^{-2}$ &\dec   5.97803768408$\times 10^{-2}$
 &\dec   5.40239556347$\times 10^{-2}$ &\dec   4.43521601235$\times 10^{-2}$ \\
  5 &\dec   1.13232334931$\times 10^{-2}$ &\dec $-$6.87114677881$\times 10^{-3}$
 &\dec $-$3.16482548625$\times 10^{-2}$ &\dec $-$6.35515201898$\times 10^{-2}$ \\
  6 &\dec   1.51540112742$\times 10^{-3}$ &\dec $-$9.22803967212$\times 10^{-3}$
 &\dec $-$2.09238359642$\times 10^{-2}$ &\dec $-$3.20631922064$\times 10^{-2}$ \\
  7 &\dec   1.04579690591$\times 10^{-2}$ &\dec   1.06631743925$\times 10^{-2}$
 &\dec   1.41218986085$\times 10^{-2}$ &\dec   2.30445918289$\times 10^{-2}$ \\
  8 &\dec   7.27178133729$\times 10^{-3}$ &\dec   4.84071702453$\times 10^{-3}$
 &\dec   1.81299605913$\times 10^{-3}$ &\dec $-$2.22929845852$\times 10^{-3}$ \\
  9 &\dec $-$2.65316666928$\times 10^{-3}$ &\dec $-$9.18394611547$\times 10^{-3}$
 &\dec $-$1.77224406625$\times 10^{-2}$ &\dec $-$2.87388172848$\times 10^{-2}$ \\
 10 &\dec $-$3.02918907022$\times 10^{-3}$ &\dec $-$3.20313568381$\times 10^{-3}$
 &\dec   4.56086296539$\times 10^{-4}$ &\dec   1.06088934260$\times 10^{-2}$ \\
 11 &\dec   4.16113147230$\times 10^{-3}$ &\dec   1.07015220405$\times 10^{-2}$
 &\dec   2.17700230213$\times 10^{-2}$ &\dec   3.73144477607$\times 10^{-2}$ \\
 12 &\dec   7.17721455928$\times 10^{-3}$ &\dec   9.49330584821$\times 10^{-3}$
 &\dec   7.35902021044$\times 10^{-3}$ &\dec $-$5.77787480240$\times 10^{-3}$ \\
 13 &\dec   3.76830635331$\times 10^{-3}$ &\dec $-$1.96870341084$\times 10^{-3}$
 &\dec $-$1.59885104470$\times 10^{-2}$ &\dec $-$4.04664282941$\times 10^{-2}$ \\
 14 &\dec   4.80548328343$\times 10^{-4}$ &\dec $-$5.66412721220$\times 10^{-3}$
 &\dec $-$1.13503013822$\times 10^{-2}$ &\dec $-$6.54745511421$\times 10^{-3}$ \\
\hline
\multicolumn{1}{c}{$i$} &\multicolumn{1}{c}{$a_i$ for $h/J_{{}_{/\!/}}=6$}
&\multicolumn{1}{c}{$a_i$ for $h/J_{{}_{/\!/}}=6.5$} 
&\multicolumn{1}{c}{$a_i$ for $h/J_{{}_{/\!/}}=7$}
&\multicolumn{1}{c}{$a_i$ for $h/J_{{}_{/\!/}}=7.5$} \\
  1 &\dec $-$3.52941176471$\times 10^{-1}$ &\dec $-$3.82352941176$\times 10^{-1}$
 &\dec $-$4.11764705882$\times 10^{-1}$ &\dec $-$4.41176470588$\times 10^{-1}$ \\
  2 &\dec $-$1.64359861592$\times 10^{-2}$ &\dec $-$1.78056516724$\times 10^{-2}$
 &\dec $-$1.91753171857$\times 10^{-2}$ &\dec $-$2.05449826990$\times 10^{-2}$ \\
  3 &\dec   1.59493430485$\times 10^{-1}$ &\dec   1.80364354971$\times 10^{-1}$
 &\dec   2.03054432723$\times 10^{-1}$ &\dec   2.27703598609$\times 10^{-1}$ \\
  4 &\dec   3.03734528767$\times 10^{-2}$ &\dec   1.16962964638$\times 10^{-2}$
 &\dec $-$1.20708465458$\times 10^{-2}$ &\dec $-$4.13195135825$\times 10^{-2}$ \\
  5 &\dec $-$1.03065008581$\times 10^{-1}$ &\dec $-$1.50607485884$\times 10^{-1}$
 &\dec $-$2.06526481612$\times 10^{-1}$ &\dec $-$2.71092352584$\times 10^{-1}$ \\
  6 &\dec $-$4.06274351241$\times 10^{-2}$ &\dec $-$4.40370259369$\times 10^{-2}$
 &\dec $-$3.91005725854$\times 10^{-2}$ &\dec $-$2.19638419584$\times 10^{-2}$ \\
  7 &\dec   4.00968938376$\times 10^{-2}$ &\dec   6.83808087160$\times 10^{-2}$
 &\dec   1.11397495384$\times 10^{-1}$ &\dec   1.72990002765$\times 10^{-1}$ \\
  8 &\dec $-$8.31253081902$\times 10^{-3}$ &\dec $-$1.83810619203$\times 10^{-2}$
 &\dec $-$3.56778569062$\times 10^{-2}$ &\dec $-$6.52026110830$\times 10^{-2}$ \\
  9 &\dec $-$4.29821115807$\times 10^{-2}$ &\dec $-$6.15590391942$\times 10^{-2}$
 &\dec $-$8.59943598221$\times 10^{-2}$ &\dec $-$1.18256118000$\times 10^{-1}$ \\
 10 &\dec   3.08422130226$\times 10^{-2}$ &\dec   6.59751313197$\times 10^{-2}$
 &\dec   1.22493770522$\times 10^{-1}$ &\dec   2.09158221300$\times 10^{-1}$ \\
 11 &\dec   5.57220816744$\times 10^{-2}$ &\dec   7.29635570762$\times 10^{-2}$
 &\dec   8.15164933911$\times 10^{-2}$ &\dec   6.90272394747$\times 10^{-2}$ \\
 12 &\dec $-$4.00161047870$\times 10^{-2}$ &\dec $-$1.09477296280$\times 10^{-1}$
 &\dec $-$2.32383122267$\times 10^{-1}$ &\dec $-$4.30523552978$\times 10^{-1}$ \\
 13 &\dec $-$7.38343538295$\times 10^{-2}$ &\dec $-$1.06305578450$\times 10^{-1}$
 &\dec $-$1.13164071654$\times 10^{-1}$ &\dec $-$4.53616541698$\times 10^{-2}$ \\
 14 &\dec   2.82846389387$\times 10^{-2}$ &\dec   1.25033040054$\times 10^{-1}$
 &\dec   3.29080311091$\times 10^{-1}$ &\dec   6.96568120714$\times 10^{-1}$ \\
\end{tabular}
\end{table}

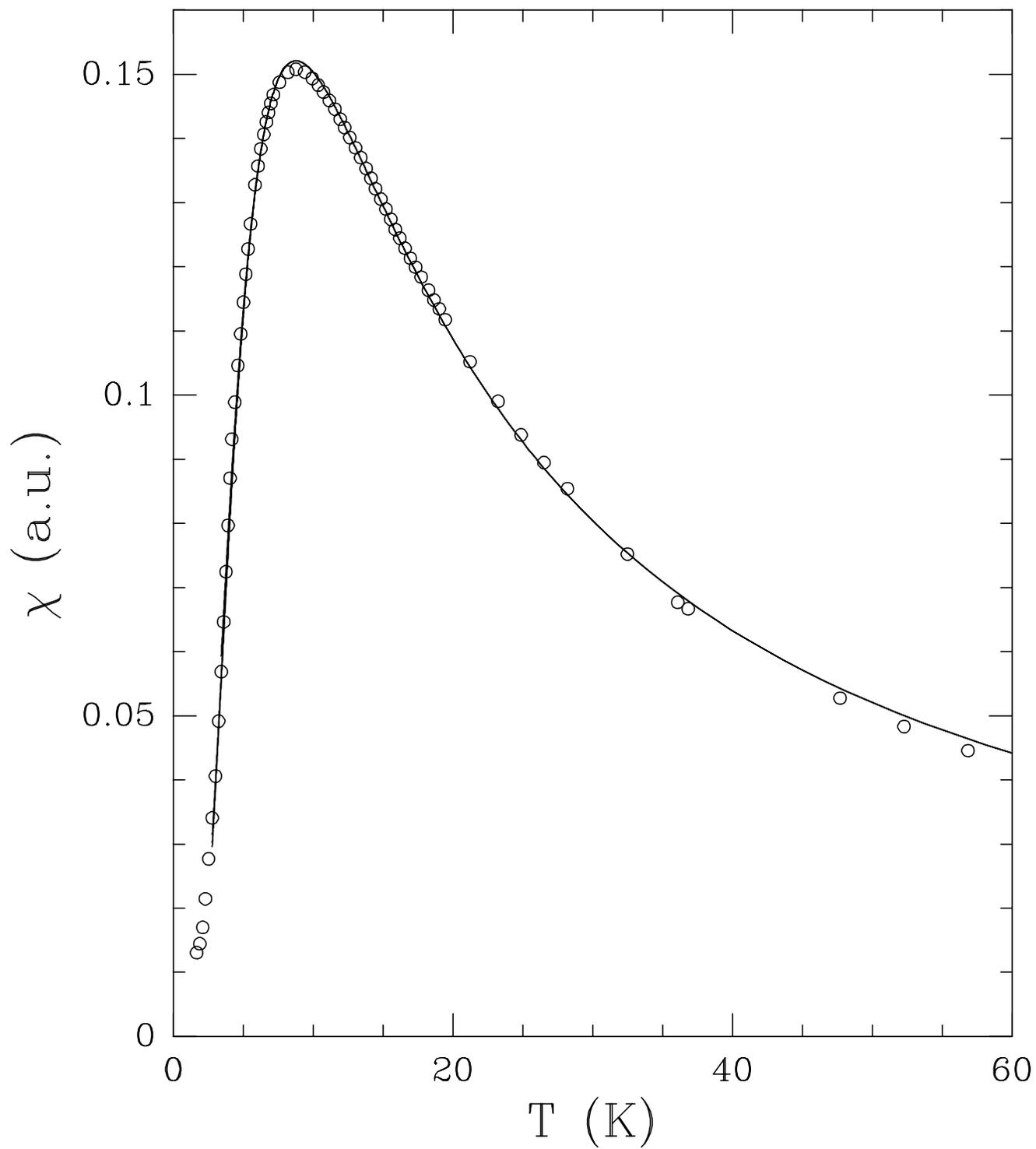
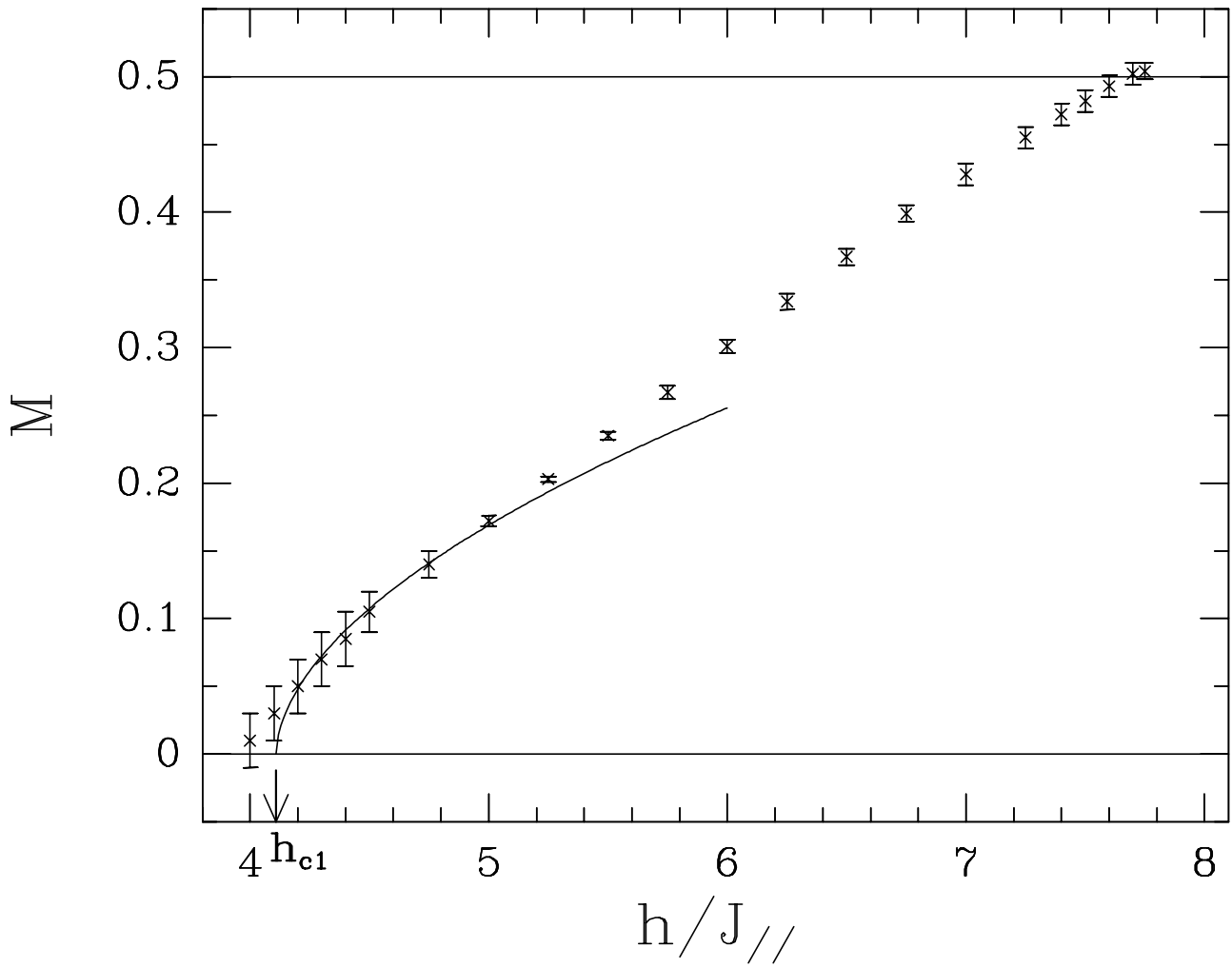
\end{document}